%Paper: gr-qc/9401010
%From: REZNIK@taunivm.tau.ac.il
%Date: Tue, 11 Jan 94 19:27:28 IST
%Date (revised): Wed, 12 Jan 94 12:49:29 IST

\input phyzzx

\def \boal {{\beta\over\eta^2}}
\def \lr {{\ln{r\over r_0}}}
\def \rh {r_0\exp{(\eta^2/\beta)}}

\def \sg { \sqrt{g^{(2)}} }

\line{\hfill TAUP-2102-93}
\line{\hfill gr-qc/9401010}
\vskip 1 true cm
 \date{}
%\vskip 1 true cm
\titlepage
\title{Entropy Generation by Tunneling in 2+1-Gravity}
\vskip 1 true cm
\centerline{ {\caps F.~Englert}$^{ab}$\footnote{*}{e-mail: fenglert@ulb.
ac.be} and
{\caps B.~Reznik}$^{b}$
\footnote{\dagger}{e-mail: reznik@taunivm.tau.ac.il} }
\vskip 1 true cm
\centerline{$^a$ Service de Physique Th\`eorique}
\centerline{Universit\`e Libre de Bruxelles, Campus Plain, C.P.225}
\centerline{Boulevard du Triomphe, B-1050 Bruxelles, Belgium}
\centerline{and}
\centerline{$^b$ School of Physics and Astronomy}
\centerline{Beverly and Raymond Sackler Faculty of Exact Sciences}
\centerline{Tel-Aviv University, Tel-Aviv 69978, Israel.}
\vskip 2 true cm
\noindent

\abstract

The tunneling approach,$^{[1]}$   for entropy generation
in quantum gravity,  is shown to be valid when applied
to 3-D general relativity.
The entropy of de Sitter and Reissner-Nordstr\"om
external event horizons and of the 3-D black hole obtained by
Ba\~nados et. al. is rederived from  tunneling of the metric
to these spacetimes.
The analysis for spacetimes with an external horizon is carried out  in
a complete analogy with the 4-D case.
However, we find
significant differences for the black hole.
In particular the initial configuration that tunnels to a 3-D black hole
may not to yield an infinitely degenerate object, as in 4-D
Schwarzschild black hole.

\vfill
\eject

\chapter{Introduction}
It is generally believed that Hawking radiation reflects
properties of quantum matter fields in the presence of an event
horizon.
Indeed in the conventional derivation of the Hawking temperature$^{
[2]}$
the spacetime is taken
to be a classical solution of Einstein's
 equations,  and the
quantization of the matter fields in this background yields a thermal
distribution with Hawking's temperature.
Then, in the standard approach, the related  entropy is
identified,  up to an additive integration constant to be $A/4$, by
using the classical laws of horizon mechanics$^{
[3]}$;
it can be associated
with our loss of information on the correlations with virtual
particles that cross the horizon.
On the other hand,
 exponentiation of the entropy should yield the number of states
of the black hole. Therefore, it was argued in the past that
the origin of this entropy, and in particular the value of the
integration constant,
could ultimately be traced only in a full theory of quantum gravity.
A clue indicating that this may be the case was provided
by the derivation of the entropy from the Euclidean
path integral approach to quantum gravity$^{
[4]}$.

A new way to obtain the entropy from a
(semi-classical) theory of quantum gravity, which presumably may yield
the value of the integration constant,
was suggested recently in Ref. $[1,5]$.
It is well known that in the absence of surface terms at infinity
there is no
``external'' time
parameter that can be used to define a unitary evolution, or a
Schr\"odinger-like equation for the matter.
However, as can be shown$^{[6]}$
, a unitary time evolution can
be restored when
the fluctuations of the metric are small. In this case, the
semi-classical approximation for gravity is appropriate, and
the spatial metric $g_{ij}$ provides a local correlation
time for the Schr\"odinger equation for the
matter.
However, in regions of strong fluctuations,
and in particular at event
horizons, this description is no longer valid and unitarity is
apparently violated. This of course, is a consequence of disregarding
the ``backward''  waves which are unavoidably generated  by the
barrier.
 It was shown, that entropy can be
regarded as measuring the departure from this apparent violation of
unitarity when a gravitational
field tunnels through a classically
 forbidden region
to  a ``classical'' region. The entropy is related to the
transmission amplitude. In this approach the
thermalization of the matter fields reflects a genuine property of
quantum gravity since it follows from the back reaction of
matter on spacetime.

The aim of this work is to analyse the approach of entropy generation
via tunneling, in 3-D general relativity.
Although in 3-D, spacetime in the absence of matter is always
flat$^{[7]}$
, coupling to extended sources leads to  spacetimes with
various topologies.
For spacetimes with a horizon, the thermodynamics of
event horizon can be  manifest with all its ramifications.$^{[8]}$.
In particular the entropy, related by the first law to the Hawking
 temperature $\kappa/2\pi$
($\kappa $ is the surface gravity) is given here by $S= L/4G$,
 where $L$ is the length of
the horizon. In any classical process the ``length'' entropy satisfies
$\delta L \ge 0$, which is the 3-dimensional analog of the second law of
thermodynamics.

In this paper we rederive this entropy  using  the
tunneling approach. We discuss separately the tunneling process for 3-D
spacetimes with an external (outer) event horizons and that to the 3-D
black hole. We show that in the case of de Sitter and  Reissner-Nordst\"om
spacetimes the tunneling process is very similar to that found in the
4-D case.$^{[1]}$
 It takes place from a wormhole, idealized as a point on the boundary
of an Euclidean
section of spacetime, to the ordinary Minkowskian spacetime.
We then analyse the tunneling to the 3-D black hole discovered recently by
Ba\~nados et. al.$^{[10]}$
 It is shown that the entropy of the 3-D black hole is
generated by the tunneling from an initial configuration of an
infinitesimally
small black hole surrounded by a matter distribution, to that of a black
hole with the same mass.
However, the initial matter configuration, differs significantly
from that found for the 4-D Schwarzschild black hole. In particular this
configuration does not seem to generate an infinitely degenerate  Hilbert space
of states.
% The entropy of a black hole is determined by the first law
%up to a integration
%constant $\cal C$ that corresponds to `internal'' states.
%In 4-D the tunneling
%approach indicates $\cal C$ could be infinite, our analyse
%suggests that in 3-D $\cal C$ is likely to vanishes, and therefore,
% the entropy of the
%black hole is given only by the ``length'' of the horizon.

The paper continues as follows. The general approach of entropy generation
by tunneling is summarized
in section 2. In section 3,
we derive the transition amplitude to the de Sitter spacetime and to
the 3-D analogue of Reissner-Nordstr\"om  spacetime where an external
horizon exists. In section 4, we study the more complicated case of
tunneling to the 3-D black hole discovered by Ba\~nados et. al.
We discuss our results in the last section.
In the Appendix we show how the singularity in  the Reissener-Nordstr\"om
spacetime, due to the point charge source, can be avoided by replacing it by a
smooth  charge-distribution with suitable mass and  pressures.
We adopt the units $G=\hbar=c=k_B=1$.

\chapter{Entropy from Tunneling}

We first review the tunneling approach$^{[1,5]}$
 and its
relation to entropy.  Consider a system of gravity coupled to source
fields with the action
$$ S_{tot} = S_{gravity} +S_{matter}. \eqn\stot  $$
Writing the metric in an ADM form
$$ds^2 = N^2dt^2 - g_{ij}(dx^i+N^idt)(dx^j+N^jdt) \eqn\ad $$
and restricting our attention to metrics with vanishing shift
functions, the action of gravity reads:
$$ S_{gravity}=
 \int (\Pi^{ij}\partial_tg_{ij} - N{{\cal H}}_G)d^3x
+\partial S, \eqn\sgrav $$
$$ \partial S = -\int {d\over dt}(g_{ij}\Pi^{ij})d^3x
-{1\over8\pi G}\int
\partial_i[(\partial_jN)g^{ij}\sqrt{g^{(2)}}]d^3x.
     \eqn\ssurf $$
$g_{ij}$ and $g^{(2)}$ are the
metric and  determinant, respectively,  of the metric on
the 2-dimensional hypersurface  labeled by the parameter
$t$. $\Pi^{ij}$, the momentum  conjugate  to the metric, is
$$ \Pi^{ij} = { \sg \over 32\pi N } (g^{in}g^{jm}
            -  g^{ij}g^{nm} ) \partial_t g_{nm} \eqn\conj $$
The Hamiltonian density for gravity is $N{{\cal H}}_G$, where
$${\cal H}_G =
         {16\pi\over\sg} \bigl( g_{in}g_{jm}\Pi^{ij}\Pi^{nm} -
                                (g_{ij}\Pi^{ij})^2 \bigr)
        -{\sg\over16\pi} R^{(2)}. \eqn\gham $$
It is assumed that the matter can (in principle) be expressed
in  a Hamiltonian form
$$S_{matter} = \int[\Pi^a\phi_a - N{{\cal H}}_M) ]d^2x.
  \eqn\smatt $$
A solution of the equation of motion must satisfy the constraint
$${{\cal H}} = {{\cal H}}_G +{{\cal H}}_M =0, \eqn\const $$
which expresses
 the invariance of the system under reparameterization
of the time coordinate.
The Wheeler-De Witt equation,
$$\hat{{\cal H}}\Psi(g_{ij},\phi_a) = 0, \eqn\wd $$
follows as one substitutes for
the momenta in \const\ the operators $-i{\delta\over\delta g_{ij}}$
and $-i{\delta\over\delta\phi_a}$.
The wave functional $\Psi(g_{ij},\phi_a)$ describes correlation
between the spatial metric (a point in superspace) and the matter
field configuration. If the fluctuations of the metric field are small
it can be described in the semi-classical limit. In this case
the metric follows a classical path in superspace and can be regarded
as a ``clock'' for the matter fields.
The clock fails if this path encounters a classically
forbidden region as indeed happens if a horizon exists.
Since horizons occur only when gravity is
coupled to matter, we split the matter fields into two parts,
$S_{matter}= S_{sc}+ S'$, where $S_{sc}$ corresponds
to the part of the matter that can be described (together with
gravity) in the semi-classical limit, while the rest of the action
$S'$
describes matter with small  fluctuations
(light particles,  radiation etc.). We shall denote by $\phi'_b$ and
${{\cal H}}'$ the fields and Hamiltonian,  respectively, of these
light degrees of freedom.
Therefore, our operational
clock consists of gravity+matter degrees of freedom for which
 fluctuations
in the ``classical'' region are small.
The clock's wave functional in the WKB approximation is
$$\Psi_c(g_{ij}, \phi_a) \simeq \exp i[\int\Pi^{ij}\partial_tg_{ij}
      d^3x + \int\Pi^a\partial_t\phi_a d^3x ].
       \eqn\wkb $$
The classical path in superspace then defines the correlation time
$\tau[g_{ij}(x),\phi_a(x)]$ which is a functional of the spatial metric
and the source fields. As long as the semi-classical limit is
satisfied the solution to the Wheeler-De Witt equation reads
$$\Psi(g_{ij},\phi_a,\phi'_b)\simeq\Psi_c(g_{ij},\phi_a)\chi(\phi'_b
         ,\tau) \eqn\wave  $$
and the wave functional for  the light matter $\chi(\phi'_b,\tau)$
satisfies the
Tomonaga-Schwinger equation with respect to the correlation time
$$i{\delta\chi\over\delta\tau(x)}= {{\cal H}}'_m \chi(\phi'_b,\tau).
         \eqn\ts $$

Consider now a path in superspace
which encounters at some point (the hypersurface
$\Sigma_1$) a classical forbidden region. This point
can be regarded as a tunneling point and the wave functional \wkb\
tunnels along a path in a Euclidean section ${\cal E}$ of superspace.
The wave functional can reenter a Minkowskian spacetime at some
point $\Sigma_2$.
Clearly, during tunneling, the unitary evolution governed
by a Schr\"odinger-like equation breaks.
In Ref. $[1]$ it was
suggest that this failure of unitarity is reflected
by a production of entropy which is ``stored'' in the clock.
The thermal nature of the matter (represented here by the fields
$\phi'_b$) follows from the back-reaction of the matter
on the clock. While this can not be demonstrated explicitly for gravity
(since the back-reaction is well beyond the scope of the
 semi-classical approximation) we would like to explain this point
at least formally. (For a non-relativistic, exact demonstration
of this idea see Ref. $[1]$).
Suppose that in the region prior to $\Sigma_1$ the total wave functional
is given by the ``forward'' wave
$$\Psi_{clock+matter}\simeq \sum_m \Psi^m_{c}\chi_m \eqn\ini $$
where $<\chi_m\mid{{\cal H}}'\mid\chi_m> = \epsilon_m $.
Than, ``after'' the tunneling, in the semi classical region beyond
the hypersurface $\Sigma_2$, the wave functional includes a
backwards moving wave (accounting for the breakdown of unitarity), and a
forward wave given by
 $$\Psi_{transmitted} \simeq \sum_m
e^{W(-\epsilon_m)} \Psi^m_c\chi_m
   \simeq \sum_m e^{W(0)+{\delta W\over\delta E_c}\epsilon_m}\Psi^m_c
 \chi_m. \eqn\final $$
where $\exp[W(-\epsilon_m)]$ is the transmission amplitude of the
clock with an energy $E_c=-\epsilon_m$.

{}From \ini\ and \final\ we conclude that
if the matter states $\chi_m$
where  distributed ``before'' tunneling with
equal probabilities, the ``final'' state \final\ indicates that the
matter fields $\phi_b'$ have a Boltzmanian energy  distribution,
and the temperature is given by
$${\delta W\over\delta E_c}
=-(T_{Hawking})^{-1} \eqn\plantem $$
$W(0)$ can be therefore, interpreted as the entropy
gained by the gravitational clock.
The entropy is given, up to an integration constant,  by
$$\ln N_0 = -2\int_{\cal E}( \Pi^{ij}\partial_tg_{ij} +
           \Pi^i\partial_t\pi_i)d^3x,   \eqn\nzero $$
where $N_0$, the inverse transmission coefficient, is the ratio of the
forward ``out wave'' at $\Sigma_2$ to the
 ``in wave'' at $\Sigma_1$.

\chapter{Tunneling to de Sitter and Reissner-Nordstr\"om Spacetimes}

Contrary to Einsteins gravity in four dimensions,
generic
classical solutions in three dimensional of gravity coupled to
an  extended source with local positive energy density
are spacetimes with an external (outer) event horizon.
If a negative cosmological constant is
 introduced, black hole like
solutions are obtained. We shall consider this case separately in
the next section.
The thermodynamical properties of spacetimes with an outer horizon
were studied in Ref. [8], where it was shown that all the
usual laws for horizon mechanics are manifested. In particular
the entropy is given up to an additive constant by  ${L/4}$,
 where $L$ is the length
of the horizon,
and the Hawking temperature by  $T={\kappa/2\pi}$ (where
$\kappa$ is the surface gravity).
In this section we rederive
this entropy.
We consider the examples of
de Sitter and Reissner-Nordstr\"om spacetimes.

The geodesically complete metric of 3-D de Sitter spacetime
in minisuperspace form is
$$ ds^2 = dt^2 - a^2 (t)\bigl(d\theta^2 + \sin^2\theta d\phi^2 \bigr),
   \eqn\minis $$
where the scale factor $a(t)$ is
$$ a(t) = \alpha\cosh{t\over\alpha} \eqn\at $$
{}From
 \conj\ , the non-vanishing gravitational momenta on the
spatial hypersurfaces, parameterized by the label $t$, are
$$ \Pi^\theta_\theta = \Pi^\phi_\phi = - {1\over16\pi}\sin\theta \
a(t) \  {da(t)\over dt}.
 \eqn\mm $$
Clearly, $\Pi^i_j$ vanishes on the $t=0$ hypersurface $\Sigma$,
which is also the surface of minimal 2-volume.
This hypersurface can be identified as a turning point of the wave
functional to an
Euclidean regime ${\cal E}$. Extrapolating ``backwards'' to the
Euclidean section, we set $t=i\tau$ and replace the scale factor \at\
by $a_e(\tau)= \alpha\cos({\tau/\alpha})$. The Euclidean section is
a 3-sphere, and the turning point above is a 2-sphere that bisects
the 3-sphere at the geometrical center. A second turning point
${\cal P}$, can be traced
on the boundary of the 3-sphere at $\tau=-\pi\alpha$.
${\cal P}$ is
a point and not a hypersurface, therefore, strictly speaking the
turning point at ${\cal P}$ should be defined as a limit of a 2-sphere
with $a \to 0$. In limit we get
 $\int \Pi^{ij}_e\partial_\tau g_{ij}d^2x \to 0 $.

We now show that the tunneling amplitude in the Euclidean
section, parameterized by the set of hypersurfaces with
$-\pi\alpha<\tau<0$, yields the correct entropy.
The cosmological constant in the Lagrangian can be viewed as playing
the role of a matter distribution of constant rest energy and a
negative pressure
$\sigma=-p_i=\Lambda$. Therefore, in the WKB expression for the
transition amplitude in eq. \nzero\ only the gravitational part
contributes.
A simple computation yields
$$ N_0 = \exp \Biggl( -{1\over4\pi} \int^0_{-\pi\alpha}d\tau \int
d\Omega^{(2)} \Bigl({da\over dt} \Bigr) \Biggr) =
e^{{\pi\over2}\alpha} \eqn\invtra $$
It can readily be verified
 that $N_0=e^S$, where $S$ is the
entropy of de Sitter spacetime. Notice that
the static metric seen by a freely falling  observer situated at a
point $r=0$ is
$$ds^2 = (1-r^2/\alpha^2)dt_s^2 - (1-r^2/\alpha^2)^{-1}dr^2 - r^2
d\theta^2.  \eqn\stat $$
The length of the (observer-dependent)  horizon is
$L=2\pi\alpha$; comparing with \invtra\ , we find that $S=L/4$.
Thus the usual  entropy can by derived
from the tunneling of what we might interpret as a wormhole
connecting to some other spacetime.

Another way to calculate $N_0$, which is useful when the
minisuperspace form of the metric is complicated,
is to use the static form \stat\ of the metric directly.
The Euclidean section obtained from \stat\ by continuation
of the open hypersurface $t_s=0$, $0<r<\alpha$ to imaginary time
$t_s=i\tau_s$, is the complete
Euclidean 3-sphere. The second turning point, the closed hypersurface
$\Sigma$ above is the union of two open surfaces parameterized by
the angular Euclidean time as $\tau_s=0$ and $\tau_s=\pi\alpha$. As
$\tau_s$ sweeps between the two values, the  hypersurface
$\tau_s=$ constant spans
exactly the same half of the 3-sphere that connects the two turning
points ${\cal P}$ and $\Sigma$ above.
Since the metric in this coordinate system is static,  $\Pi^i_j$
vanishes along the hypersurfaces $\tau_s=$ constant.
Using this property and the covariance of the action \sgrav\ ,
we can express
the inverse transition coefficient as a surface term
$$ N_0 = \exp\bigl(-2S_e(\Sigma,{\cal P}) \bigr) =
        \exp\Biggl( -
{T_e\over4\pi}\int\partial_i[(\partial_jN)g^{ij}\sqrt{ g^{(2)} } ]
d^2x \Biggr) \eqn\nsur  $$
It can be easily  verified that \nsur\ with the metric \stat\ yields
the same result as in \invtra\ .

Next we consider the 3-D spacetime generated by a charged
source.
The static solution of  Einstein-Maxwell equations, corresponding to
Reissner-Nordstr\"om \hfill\break
spacetime is
$$ds^2 = (1-\boal\lr)dt^2 - \eta^{-2}\biggl({1-\boal\lr}\biggr)dr^2
          -r^2d\theta^2, \eqn\rn $$
where $\beta=4Q_0^2$, $\eta=1-4m$ and $Q_0$ and $m$ are the
charge and mass of the source, respectively. \rn\ possesses
a regular singularity at $r=r_h\equiv\rh$ that corresponds to
an outer event horizon. A second curvature singularity ($
R_{ab}R^{ab}=\beta^2/r^4$) exists
at $r=0$ due to the divergence of the electric field at this point.

A Kruskal extension of \rn\ yields a geodesically complete
spacetime with
topology  $S^{(2)}\times
R^{(1)}$. Any  Cauchy hypersurface contains two point charges of opposite
sign, which are located on ``antipodal'' points.
Analytically continuating \rn\  to an Euclidean time
$\tau=-it$, we find that the hypersurfaces parameterized by $\tau$
span a compact Euclidean section of a topology $S^{(3)}$ with
a period $T_e= 4\pi r_0 {\eta\over\beta}\exp(\eta^2/\beta)$.
It can be readily verified that the
open hypersurface $t=0$, $0<r<r_h$ in \rn\ is one half of a closed
hypersurface $\Sigma$. The second half of
$\Sigma$ is the $\tau=
T_e/2$. Clearly, all the gravitational momenta vanish on $\Sigma$, and
this hypersurface is a turning point to an Euclidean section of
spacetime.
The other turning point ${\cal P}$ is located as in the de Sitter
spacetime on the boundary of the Euclidean 3-sphere, as is shown
below.

Since it is much more complicated in this case to construct a metric
with a  minisuperspace form we will
compute the inverse transition coefficient
from the surface term \nsur\ .
However we cannot use
the metric \rn\ , since
continuation of the static metric \rn\ yields
an  Euclidean spacetime with a singularity at
%A simple
%computation yield $(\partial_jN)g^{ij}\sqrt{ g^{(2)}}= constant$ and
%therefore,  $\ln N = 0$. Further reflection shows that the difficulty
$r=0$
on the boundary.
 This singularity must be
removed, either by a suitable regularization of the metric near the
origin, or
by replacing the point charge source by a smooth charge distribution
that yield a regular metric at $r=0$.
The latter possibility of replacing the source by
a smooth charge-density distribution,
with suitable pressures that keep the system static, is studied in
the Appendix.
In general, it is sufficient  that
the metric near the
origin behave as $g^{rr}\simeq g_{tt} \simeq 1 + Cr^\delta$
with $\delta>0$.
In this case, the contribution of the surface term at $r=0$
vanishes and the inverse transmission coefficient  is given by
$$\ln N_0=
-2\pi{T_e\over8\pi}\biggl((\partial_rN)g^{rr}\sqrt{g^{(2)}}
           \biggr)^{r=r_h}_{r=0}={\pi\over2}r_0\exp{\eta^2\over\beta}
 =  {L_{RN}\over4}.       \eqn\entropy $$

Under the latter condition of a smooth metric
near the origin we can indeed verify  the existence in the limit of the
turning point ${\cal P}$. In the coordinates $T$, and $\xi$ defined by
$$ r= T\sin\xi, \ \ \ \ \tau - \tau_0= T\cos\xi. \eqn\tran   $$
the Euclidean continuation of the metric \rn\ near $r \to 0$
yields
$$ ds^2 \simeq -dT^2 -T^2(d\xi^2 + \sin^2\xi d\theta^2) .\eqn\appmet
          $$
The gravitational conjugate momenta on the limiting
2-sphere  vanish as $T \to 0$ . Therefore, any point ${\cal P}$ ($r=0,
\tau=\tau_0$)
is a second turning point.
 The Lorenzian Reissner-Nordstr\"om spacetime with its
two image charges can be viewed as emerging via tunneling through the
Euclidean section from a wormhole, idealized
here as the point ${\cal P}$.

We have studied in details only de Sitter and
Reissner-Nordstr\"m spacetimes. With regularized sources
our
result is valid also to general solutions of 3-D gravity with an
outer horizons, as clearly can be seen from the analysis above.

\chapter{Tunneling Entropy of a Black Hole}

The topology of the Euclidean black hole in $n$ dimensions is
$R^{(2)}\times S^{(n-2)}$.
If the topology is unchanged by  tunneling
one cannot expect the tunneling to occur from
a wormhole-like turning hypersurface as in  the case of a closed
``universe'' with $S^{(n)}$ Euclidean topology considered
in the previous section.
Indeed, the scenario suggested for the
Schwarzschild black hole
is that the tunneling occurs from
a static configuration of a
``germ'' black hole surrounded by ordinary matter to a static
black hole
of total equal mass.
 The germ black hole of mass $m_0$ is surrounded by a concentric
static thin shell of matter with mass  $m-m_0$ and radius
$r_s$, which is
located outside the Schwarzschild radius of the total mass $m$.
In the limit of a thin shell that is located
infinitesimally close to $2m$ and when $m_0 \to 0$,
this configuration is connected by a Euclidean section of spacetime
to that of a black hole of the same total mass $m$, and the tunneling
between the two Minkowskian sections was shown to generate the entropy
of the black hole.

In the following we study this problem for the 3-D
black hole obtained recently by Ba\~nados $et. al.$$^{[10]}$.
When
a negative cosmological constant is introduced, the Einstein equations
admit a black hole solution (G=1)\footnote{\dagger}{Other definitions
for $G$ are sometimes used in the literature depending on the choice
of the coefficient in the action. We use $1/16\pi G\int R$.}
$$ds^2 = \Bigl(\Lambda r^2 - 8m_0 \Bigr) dt^2 -
\Bigl(\Lambda r^2 - 8m_0 \Bigr)^{-1}dr^2 - r^2d\theta^2. \eqn\bhmet $$
$\Lambda >0$ is the negative of the cosmological constant,
and $m_0$ the mass. The entropy and Hawking temperature of the black
hole, derived the standard way, are respectively
 $S= {\pi\over2}r_h = L/4$ ($r_h =
\sqrt{8m_0/\Lambda}$), and $T=\Lambda r_h/2\pi$.
Notice that in contrast
 to the 4-dimensional black hole, the temperature
vanishes as the black hole evaporates.

Consider a black hole of mass $m_0$ surrounded by a spherically
symmetric static matter distribution characterized by the density
function $\sigma(r)$, and the pressure functions  $p_r(r)$ and
$p_\theta(r)$. In the static coordinate system
$$ds^2 = g_{tt}(r) dt^2 - g_{rr}(r)dr^2 - r^2 d \theta^2,
\eqn\gen $$
the metric is  entirely determined by $\sigma(r)$ and $p_r(r)$.
$$g^{rr}(r)  = \Lambda r^2 -8m(r),  \eqn\grr $$
and
$$g_{tt}(r) = \Bigl(\Lambda R^2-8m \Bigr) \exp\biggl(
               -\int_r^R
{2r'(8\pi p_r +\Lambda)\over {\Lambda r'^2 -8m(r')}} dr'
\biggr),      \eqn\gtt $$
where
$$m(r) = m_0 +  \int^r_{r_h} 2\pi r'\sigma dr'. \eqn\mr $$
We have assumed that the shell is located between the radius
$r_s$
and $r_s +d $, and that the point $R$ is located outside the shell,
($R>r_s+d$).
Outside the shell ($r>r_s+d$) this metric coincides
with that of a black hole with total mass $m$.
On the other hand in the interior of the shell ($r_h<r<r_s$)  we
have
$$ g^{int}_{tt} = (\Lambda r^2 - 8m_0)
    {\Lambda(r_s+d)^2-8m\over \Lambda r_s^2 - 8m_0 }
\exp \biggl(- \int_{r_s}^{r_s+d}
{2r'(8\pi p_r +\Lambda)\over {\Lambda r'^2 -8m(r')}} dr' \biggr).
 \eqn\gttapp
$$
The angular pressure, $p_\theta$, is determined by the
equation
$$p_\theta = r^2(\sigma + p_r){8\pi p_r + \Lambda \over \Lambda r^2 -
            8 m(r) } + p_r + r {dp_r\over dr}, \eqn\equ $$
which is the 2+1-dimensional analog of the Tolman-Oppenheimer-Volkoff
equation.

We shall consider
first the possibility of constructing, as in the 4-D case,
a static thin shell
infinitesimally close  to the horizon $r_h(m)$ corresponding to the
total mass $m-m_0$ of the shell and $m_0$ of the mini black hole.
We let $r_s = r_h(m) + \epsilon$ and take the limit  $\epsilon\to
0$ and $d \to 0$.
By eq. \equ\ in the thin shell limit $p_\theta$ diverges and therefore
the ``dominant energy condition''$^{[11]}$
 is violated. This difficulty, however,
 can by
avoided by requiring that inside the shell the radial pressure is
determined by %
$$ 8\pi p_r + \Lambda = 0 \eqn\negp $$

A necessary condition for tunneling from
the shell configuration to a black hole of mass $m$,
is that
the Euclidean periodicity, and
 therefore the global temperature,
 of the two spacetimes be  identical.
The global temperature of the metric \gen\ for a mini-black-hole
with a shell is
$$ T(m_0,m-m_0) = {\Lambda r_h(m_0)\over2\pi} \sqrt
    {\Lambda(r_s+d)^2-8m\over \Lambda r_s^2 - 8m_0 }
\exp\biggl(-
\int_{r_s}^{r_s+d}
{r'(8\pi p_r +\Lambda)\over {\Lambda r'^2 -8m(r')}} dr' \biggr)
\eqn\ttt $$
In the limit $m_0 \to 0$, $T(m_0)=\Lambda r_h(m_0)/2\pi \to 0$, by
\negp\ the exponential factor gives 1, and
therefore, $T(m_0, m-m_0) \to 0$.
We see that the situation is opposite to that in 4-D. There, the
temperature of the germ black hole  diverges as $m_0 \to 0$ but the
shell screens this temperature. We conclude that the thin shell
configuration is inappropriate to a 3-D black hole.

The difficulties we have encountered
can be resolved by replacing the thin shell by another initial
configuration.
Consider, instead, a shell of the same mass $m-m_0$, which
extends from $r_2 = r_h(m)+\epsilon$ to $r_1 = r_h(m_0)+\xi$. The matter
is distributed on a ``thick'' shell starting at
 the horizon $r_h(m_0
) $ of the germ black
hole and ending at an infinitesimal distance outside the location
of the horizon $r_h(m)$ of a black hole of mass $m$.
The whole system is on the
verge of forming a black hole but, as before it is static.
Requiring that $p_r$ inside the shell satisfy Eq. \negp\ we
get the global temperature
$$T(m_0,m-m_0, \epsilon,\xi) =
 {\Lambda r_h(m_0)\over2\pi} \sqrt
    {\Lambda(r_h(m)+\epsilon)^2
    -8m\over \Lambda (r_h(m_0)+\xi)^2 - 8m_0 }.\eqn\ttwo
$$
Clearly, as $m_0,\epsilon,\xi \to 0$ by a suitable limiting process
we may obtain
$$ \lim_{m_0,\epsilon,\xi \to 0} T(m_0,m-m_0,\epsilon,\xi)
= T(m).\eqn\tl $$
Therefore, the Euclidean period of the configuration is the same as
that of the Euclidean period of a black hole of mass $m$, and the
two Euclidean spacetimes geometries coincide for $r>r_h(m)+\epsilon$.

%By \negp\ and \equ\ we have $p_r = p_{\theta} =
% - \Lambda/8\pi$.

 In the limit $m_0 \to 0$ we get by \mr\
 that the only possible configuration is that of a uniform matter
 density $\sigma= |p_r| = |p_\theta|$. The ``dominant energy condition''
 $T^{00} \ge  |T^{ii}|$
is satisfied inside the shell.\footnote{\dagger}{ The discontinuity at
 the boundaries of the shell can be avoided by
letting the densities die continuously
to zero while keeping  $\sigma + p_r= 0 $.}
Any  non-uniform
matter distribution with the same mass extension will not satisfy
the dominant energy condition; relaxing equation \negp\ will not
help.

Notice that the matter configuration we have found amounts
to adding a positive cosmological constant to the region bound by
$r(m)$ and $r(m_0)$.
 If we interperete the negative cosmological constant
as a source, we have in this region a ``superposition'' two sources,
corresponding to a negative and a positive cosmological constants,
which sum up to zero. Therefore, in the limit we use,
the sources vanish the shell.
and spacetime is flat. However, inside the shell we have
$$ g_{tt} = \Lambda(r_h(m)+\epsilon)^2 - 8m \sim\Lambda r_h(m)\epsilon
   \to 0.\eqn\gttz $$
There is an infinite
red shift with respect to an observer outside the shell.

We now identify  the first turning point
$\Sigma_1$ as the union of the $t=0$
hypersurface
of \gen\ and the other static half in the Kruskal extension.
The second turning point, back to the Minkowskian spacetime,
 is the $t=0$ hypersurface $\Sigma_2$ of the
Kruskal metric of a 3-D black hole  with mass $m$.
In the limit of $m_0,\epsilon,\xi \to 0 $ the two turning points
are connected by a Euclidean section of spacetime. This can be seen by
nothing that the Euclidean continuation of $\Sigma_1$ and $\Sigma_2$
matches for $r>r_h(m) +\epsilon$, while the 3-volume $V_{\cal E}$
of the region
$r_h(m_0)<r<r_h(m)$ vanishes in this limit, due to the vanishing
of $g_{tt}$ inside the shell.
 The Euclidean region through which the wave function
tunnels is, therefore, the region  $\cal E$.
Although the volume of this region vanishes,
we will readily show that
the Euclidean action between $\Sigma_1$ and $\Sigma_2$, which gets
contributions only  from $\cal E$, does not vanish.
The inverse transmission amplitude can be expressed
by the difference of the Euclidean actions of the black hole and
the germ black hole with the  shell

$$ N_0 = \exp \Bigl( S_e(\Sigma_1) - S_e(\Sigma_2) \Bigr). \eqn\sms $$
Since in this static gauge
the gravitational and matter momenta vanish everywhere, the
actions are given by the last surface term in Eq. \ssurf\ .
This yields
$$  N_0 =  \exp \Bigl( \pi r_h(m)/2  -\pi r_h(m_0)/2
\Bigr).\eqn\nbh   $$
In the limit $m_0 \to 0$ we get
$$N_0 = \exp {L\over4}.\eqn\bhen $$

\chapter {Conclusions}

The entropy associated with external horizons and the 3-D black hole has
been rederived from the tunneling approach to entropy generation.
The analysis for external horizons is similar to the 4-D case. We found
however,
important differences in the case of a black hole.

Although the entropy of the 3-D black hole was correctly derived by
this approach,
the initial configuration that tunnels to the black hole is
more constrained than in 4-D.
Classical considerations seem to indicate that
this object has
small or vanishing degeneracy in contrast to the infinite degenerate
object found
in the case of tunneling to the Schwarzschild black hole.$^{[5]}$
This should be
verified explicitly
by including also quantum fluctuation of the shell.
Yet, if the tunneling approach for the eternal black
hole does hint on the dimensionality of
the Hilbert space of the collapsing
black hole, and if indeed the initial configuration
that tunnels to the black hole is not degenerate,
then this
indicates a significant difference between 3-D and 4-D black holes.
It suggests that for the 3-D  black hole only
the ``length''  entropy is relevant and the possible
 additive constant is vanishing or small.
 This  seems quit reasonable, since the  3-D black hole
 evaporates
in  an infinite time, and if proper boundary condition are taken at
infinity the decay may even eventually stop.

\vskip 2 true cm

\ack
The research was supported in part by grant 425-91-1 of the
Basic Research Foundation, administered by the Israel Academy
of Sciences and Humanities .

\vfill\eject

\chapter{Appendix}

We shall verify that under suitable conditions the point charge source
of the 3-D Reissner-Nordstr\"om can be replaced by a smooth charge
density such that the metric becomes regular at the origin. To this
end consider a spherically symmetric charged perfect fluid coupled to
Einstein-Maxwell equations:
 $$ G_{ab} =8\pi G (T^f_{ab} + T^{em}_{ab}),
\eqn\ein $$
 $$\nabla_b F^{ab} = J^a,$$
$$\nabla_{[a}F_{bc]} =0.$$
 The source is the sum of
 $$T^f_{ab}= \rho u_a
u_b + P(u_a  u_b -g_{ab}) \eqn\femt $$
 and
 $$
T^{em}_{ab}=F_{ac}{F^c}_b + {1\over4}
g_{ab}F^2. \eqn\ememt $$
Let
 $$ds^2 = g_{tt}(r)dt^2 -
g_{rr}(r)dr^2 -r^2d\theta^2,\eqn\metric $$
 $$F_{10} = E(r).$$
 The  solution of
Maxwell equations is
 $$E(r) = \sqrt{g_{tt}(r)g_{rr}(r)}
{Q(r)\over2\pi r},\eqn\elec $$
 where
 $$Q(r)=\int^r_0 2\pi r \sigma(r)dr. \eqn\charge $$

{}From Einstein's equations we derive
% $$R_{00}=\half
%e^{\xi-\psi}(\xi''+\half\xi'^2-\half\xi'\psi'+ {\xi'\over r})=16\pi
%Gp(r)e^{\xi},\eqn\rzero $$
%%
%$$R_{11}=-\half(\xi''+\half\xi'^2-\half\xi'\psi'-{\psi'\over r})=
%8\pi G(\rho-p)e^{\psi},\eqn\rone $$
% $$R_{22}=-\half re^{-\psi}(\xi'-\psi')=
%8\pi G(\rho-p)r^2 +{2G\over\pi}Q^2(r). \eqn\rtwo $$
% By equations
%\rzero\ and \rone\
% $$\half e^{-\psi}(\xi' + \psi') = 8\pi G(\rho +p)
%\eqn\plus $$
% which leads together with \rtwo\ to the relations
%$$e^{-\psi}\psi' = 16\pi G\rho r +{2G\over\pi}{Q^2\over r} \eqn\one
%$$
%$$e^{-\psi}\xi' =16\pi Gprr - {2G\over\pi} {Q^2\over r}. \eqn\two $$
%Also since $e^{-\psi}R_{11}-{1\over r^2}R_{22} = -{2G\over\pi}
%{Q^2\over r}$ we get the relation
% $$\xi''+\half
%\xi'^2-\half\xi'\psi'-{\xi'\over r}= {4G\over\pi}e^{\psi}{Q^2\over
%r^2} \eqn\pure $$
%
% Integrating equations \one\ and \two\ we get
%
%
$$g_{rr}^{-1}  = 1 - m (r) \eqn\grr $$
 and
 $$g_{tt}(r_2) - g_{tt}(r_1) = \int^{r_2}_{r_1} \exp \biggl(16\pi
G{p(r)r - {1\over8\pi^2}{Q^2\over r}\over 1-m(r)} \biggr),
\eqn\gtt $$
 where
 $$m(r) =
16 \pi G \int^r_0 (\rho r + {1\over8\pi^2} {Q^2\over r} )dr.\eqn\mdef
$$

To complete the solution we need the equation of state of the fluid,
that is a relation $p=p(\rho, Q)$. We find
%Differentiating equation \two\ and
%substituting $\xi''$ from \pure\ leads to
% $$p' = -{1\over32\pi
%G}\xi'{e^{-\psi}\over r}(\xi'+\psi') +{1\over8\pi^2}{{Q^2}'\over r^2}
%$$
% utilizing \plus\ and \xiprime\ gives finally the relation
 %
%
 $$p' =
8\pi G(\rho +p){{1\over8\pi^2}{Q^2\over r} -pr\over1-m(r)}
+{1\over8\pi^2}{{Q^2}'\over r^2}.\eqn\state $$
 To the first order we
can see that equilibrium is obtained as a negative pressure balances
the ``gravitational'' repulsion $G\rho{Q^2\over r}$ and the electric
repulsion ${{Q^2}'\over r^2}$.

Clearly, if the charge density vanishes at the origin fast enough, by
\gtt\ \grr\ and \mdef\ the metric will be regular at the origin. In
fact these equations can be integrated for special cases.
Assuming that the charge distribution is given by
$Q(r)=Q_0({r\over r_0})^n$ and that $p=-\rho$ (the dominant energy
condition holds) the fluid will
remain in a state of equilibrium provided that the equation \state\ is
satisfied.  Therefore,
 $$p' = {1\over8\pi^2}{{Q^2}'\over r^2}. \eqn\psp
$$
 When $n\leq 1$ the pressure and $g_{rr}$ diverge at the
origin.  Therefore consider only the case
$n> 1$.  Integrating \psp\
 $$\rho (r) = {1\over8\pi^2}{n\over
n-1}({Q_0\over r_0})^2 (1- ({r\over r_0})^{2n-2}).\eqn\denpres $$
 The
pressure and the density vanish at the at the surface of the
``star'' at $r=r_0$.  The integration of equations \grr\ leads to
$$g^{(int)}_{tt}=-g^{rr}= 1-{n\over n-1}{GQ_0^2\over\pi}({r\over
r_0})^2 (1-{1\over n^2}({r\over r_0})^{2n-2}). \eqn\intmet $$
 %
% where
%we have utilized also equation \one\ .
The electric field inside the ``star'' is
 $$ E(r) = ({Q_0\over r_0)}^2 ({r\over
r_0})^{2n-1}.\eqn\ele $$
 The exterior metric ($r> r_0$) obtained by integrating
\grr\ with $Q(r)=Q_0$ and $\rho=p=0$ is
$$g^{(ext)}_{tt}=(g_{rr}^{(ext)})^{-1}= 1- {n+1\over
n}{GQ_0^2\over\pi}- {2\over\pi}GQ_0^2\ln {r\over r_0}.\eqn\extmet $$
Finally, the total mass of the star is given by
 $$M= 2\pi\int^{r_0}_0
T_0^0\sqrt{g^{(2)}} dr =-{1\over4G}(\sqrt{ 1-{n+1\over
n}{GQ_0^2\over\pi}} -1). \eqn\mas $$
 Therefore, the exterior metric can be written
as
 $$g_{tt} = \eta^2(1- {\beta\over\eta^2}\ln {r\over r_0}),\eqn\extm $$
where
 $$\eta = 1-4GM, \ \ \  \beta= {2GQ_0^2\over\pi}\eqn\consn $$
 are related to
the total mass and total charge of the star.

\vfill \eject

{\bf REFERENCES}

\Item{1.} A.~Casher and F.~Englert, Class.  Quantum Grav., {\bf 9},
          2231 (1992).
\Item{2.} S.~W.~Hawking, Commun. Math. Phys., {\bf 43}, 199 (1975).
\Item{3.} J.M.~Bardeen, B.~Carter and S.W.~Hawking,
          Cummun. Math. Phys. {\bf31}, 161 (1973).
\Item{4.} S.~W.~Hawking, {\it The path-integral approach to quantum
          gravity}, published in {\it General Relativity. An Einstein
          centenary survey}, ed. S. W. Hawking and W. Israel,
           Cambridge University
          Press, 1979.  And references within.
\Item{5.} A.~Casher and F.~Englert, {\it Black Hole Tunneling
          Entropy and the Spectrum of Gravity}, ULB-TH 8/92,
          TAUP-2017-92, to be published in Class. Quant. Grav.
\Item{}  A.~Casher and F.~Englert, {\it Entropy Generation in
         Quantum Gravity and Black Hole Remnants},
         {\it in} Proceeding of the Int. Conf. on Fundamental Aspects of
         Quantum Theory, to celebrate the 60th birthday of Yakir Aharonov,
         December 10-12, 1992, Columbia, SC, to be pub.
\Item{6. } V.~G.~Lapchinsky and V.~A.~Rubakov, Acta.  Phys.  Pol. {\bf
         B10}, 1041 (1979).
\Item{}  T.~Banks, Nucl.  Phys. {\bf 249}, 332 (1985).
\Item{}  R.~Brout, Foundation of Phys. {\bf 17}, 603 (1987).
%\Item {} R.~Brout, G.~Horwitz and D.~ Weil, Phys.  Lett. {\bf B192},
% 318 (1987).  R.~Brout
         R.~Brout and G.~Venturi, Phys.  Rev. {\bf D39}, 2436 (1989).
\Item{7.} A.~Staruszkiewicz, Acta. Phys. Polon. {\bf 24}, 739 (1963).
\Item{}   S.~Deser, R.~Jackiw and G.~t'Hooft, Ann. Phys., {\bf152}, 220 (1984).
\Item{8.} B.~Reznik, Phys. Rev. D {\bf 45}, 2151 (1992).
\Item{9.} S.~Deser and P.O.~Mazur, Class. Quan. Grav. {\bf2}, L51
          (1985).
\Item{}   J.R.~Gott, III, J.Z.~Simon and M.~Alpert, Gen. Rel. Gra.
          {\bf18},  1019 (1985).
\Item{}   B.~Reznik, ``{\it Charged Point Sources in 2+1-Dimensional
          Gravity}'', TAUP-1834-90, unpublished.
\Item {10.}  M. Ba\~nados, C. Teitelboim and J. Zanelli, Phys. Rev.
             Lett., {\bf 69}, 1849 (1992).
\Item {}  M. Ba\~nados, M.~Henneaux, C. Teitelboim and J. Zanelli,
          Phys. Rev. D {\bf 48}, 1506 (1993).
\Item {11.} S.~W.~ Hawking and G.~F.~R.~ Ellis, {\it The Large Scale
           of Space-Time} , Cambridge University Press, Cambridge,
            England, 1973.

\vfill
\eject

\bye